\documentclass[reprint,unsortedaddress,amsmath,amssymb,aps,pra,showpacs]{revtex4-2}
\usepackage{amsmath}%
\usepackage{amsfonts}%
\usepackage{amssymb}%
\usepackage[cal=cm]{mathalfa}
\usepackage{graphicx}
\usepackage{dcolumn}
\usepackage{bm}
\usepackage{braket}
\usepackage{sidecap}
\usepackage{color}
\usepackage{bbm}
\usepackage{nicefrac}
\usepackage{float}
\usepackage{placeins}
\usepackage[dvipsnames]{xcolor}
\usepackage[utf8]{inputenc}
\usepackage[normalem]{ulem}
\usepackage{nicefrac}
\usepackage{dsfont}
\usepackage{xcolor, soul} 
\usepackage[normalem]{ulem} 
\usepackage{siunitx}
\usepackage[citebordercolor=blue]{hyperref}
\usepackage[utf8]{inputenc}
\usepackage{gensymb}
\usepackage{mathtools}

\begin{document}

\title{Orbital Angular Momentum Generation in Neutron Schwinger Scattering from Perfect Quartz}

\author{Niels Geerits$^{1}$}
\email{niels.geerits@tuwien.ac.at}
\author{Anna-Sophie Berger$^1$}
\author{Hartmut Lemmel$^{1,2}$}
\author{Steven R Parnell$^{3,4}$}
\author{Jeroen Plomp$^3$}
\author{Michel A. Thijs$^3$}
\author{Stephan Sponar$^1$}
\email{stephan.sponar@tuwien.ac.at}
\affiliation{%
$^1$Atominstitut, Technische Universit\"at Wien, Stadionallee 2, 1020 Vienna, Austria \\ $^2$Institut Laue-Langevin, 71 Avenue des Martyrs, CS 20156, 38042
Grenoble Cedex 9, France \\
$^3$Faculty of Applied Sciences, Delft University of Technology, Mekelweg 15, Delft 2629JB, The Netherlands \\
$^4$ISIS, Rutherford Appleton Laboratory, Chilton, Oxfordshire, OX11 0QX, UK}

\date{\today}
\hyphenpenalty=800\relax
\exhyphenpenalty=800\relax
\sloppy
\setlength{\parindent}{0pt}
\begin{abstract} 
Static electric fields have been suggested as a spin to orbital angular momentum converter in neutrons. Initial calculations showed that the field required to facilitate significant conversion to longitudinal orbital angular momentum is prohibitively high for lab power supplies. In this work we exploit the intra-atomic nuclear electric field in the periodic structure of perfect single crystals, specifically quartz, which can be orders of magnitude larger than lab fields. We calculate the Bragg and Laue diffracted wavefunctions of thermal neutrons and back-diffracted neutrons and demonstrate spin to orbital angular momentum conversion. Finally we report on a thermal neutron Bragg diffraction experiment from [110] quartz confirming our results.
\end{abstract}

\maketitle
\section{Introduction}
Neutron Orbital Angular Momentum (OAM) has been receiving increasing interest in literature \cite{Clark2015,Nsofini2016,Sarenac2018,Sarenac2019,Ivanov2022,Sarenac2022,Thien2023,Geerits2023}, as it is expected that neutrons carrying OAM could interact differently with matter \cite{Afanasev2019,Afanasev2021,Jach2022}, such that twisted neutron scattering could reveal additional information on the properties of nuclei and atomic and crystal structure. Also, as an additional quantum mechanical degree of freedom neutron OAM could be of interest for entangled measuring probes or quantum contextuality experiments \cite{Shen2020,Kuhn2021}. Though the generation of intrinsic neutron OAM remains challenging. In the first instance, described in 2015 \cite{Clark2015}, an aluminum spiral phase plate was used to generate a phase vortex across the macroscopic beam cross section. This was later critiqued in 2018 \cite{Cappelletti2018}, since it could be demonstrated that the neutrons in such a beam carry only extrinsic OAM relative to the optical axis of the spiral phase plate. OAM is considered to be intrinsic when it is invariant under translation of the frame of reference in which it is measured \cite{Berry1998,Neil2002,Geerits2023}. This occurs when the expectation values of the transverse momentum components are zero. Since the 2015 experiment significant progress has been made. Most notably a 2022 experiment \cite{Sarenac2022}, which reports on an array of microscopic fork gratings, each roughly the size of the transverse neutron coherence length (in this case $1\mathrm{\mu m}$). Here the grating scatters neutrons into three sub-beams, a direct beam carrying no OAM and two sub-beams mirrored around the central beam carrying opposite OAM. Though as the transverse momentum of the sub-beams is non zero, the OAM of the individual sub-beams cannot be considered strictly intrinsic (cf. Sec II). In addition, the production of such grating arrays is for the time being expensive, hence a more accessible technique may be of benefit. Finally, with current techniques, effective gratings of this type can only be produced for cold neutron beams. Hence the challenge remains to develop a, preferably simple, method for OAM generation in thermal neurons. In the mean time, work has continued on developing other methods to generate neutron OAM, mostly focused on coherent averaging \cite{Sarenac2018,Sarenac2018b,Sarenac2019,Geerits2023}. However in recent years we have explored the possibility of using strong electric fields, which couple neutron spin to momentum, to twist neutron wavefronts \cite{Geerits2021} by the Schwinger interaction \cite{Schwinger1948}. While electric fields generated in the lab are too weak to meaningfully affect the neutron, the naturally occurring fields in atoms are of a sufficient magnitude to generate OAM states in neutrons. Fields of up to $10^{10} \frac{\rm V}{\rm m}$ have been exploited in neutron diffraction from perfect quartz \cite{Voronin2000}. In this paper we report on an experiment employing reflection off the 110 plane of perfect quartz in Bragg geometry to generate neutrons in an OAM state, more precisely a spin-orbit state, where the spin and OAM degree of freedom are entangled \cite{Nsofini2016}. In addition using the dynamical theory of diffraction \cite{Batterman1964,Sears1978}, we provide a theoretical model for the wavefunction of a neutron that is reflected/transmitted from/through the 110 plane of Quartz in both Laue and Bragg geometry. Using this model we calculate the expected intensity and polarization profile, in addition to the OAM distribution.
First, however, we will provide a short introduction to OAM and introduce the toolbox used to analyze the OAM of neutron wavefunctions. 
\section{Orbital Angular Momentum}
\begin{figure}[!h]
	\centering	\includegraphics[width=8cm]{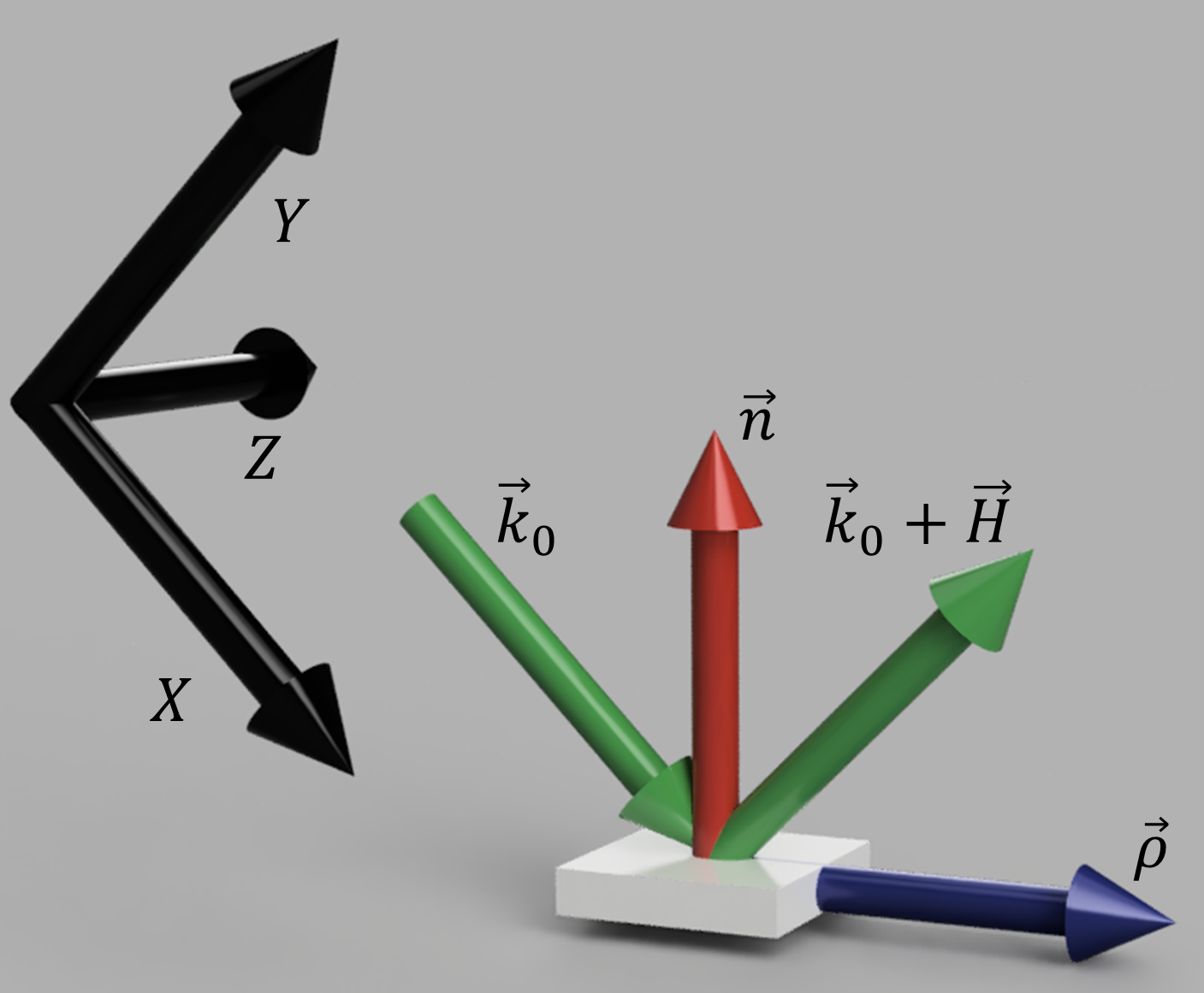}\caption{Three dimensional render depicting Bragg diffraction of a neutron (green) by a crystal (white). The axes that define the spin operators $\mathbf{\sigma}$ are shown in black, the surface normal $\mathbf{n}$ in red and the so called $\rho$ axis in blue. Rocking represents a rotation of the surface normal towards or away from $\rho$, while a rotation around the $\rho$ axis is defined as tilting.}
	\label{CrystalAxes}
\end{figure}
In quantum mechanics OAM is defined identically to angular momentum in classical physics, 
\begin{equation}
    \hat{\mathbf{L}}=\hat{\mathbf{r}}\times\hat{\mathbf{p}}
\end{equation}
except the position, $\mathbf{r}$, and momentum, $\mathbf{p}$, vectors are upgraded to operators in the quantum mechanical case. The Eigenstates of the $k^{\mathrm{th}}$ component of the OAM operator are given by $e^{\mathrm{i}\ell\phi}$ with $\phi$ the azimuthal coordinate defined in the plane orthogonal to the direction of the $k^{\mathrm{th}}$ component. Given that these Eigenfunctions form a complete basis we can expand any azimuthal structure in terms of its OAM modes. To this end we introduce the azimuthal Fourier transform (AFT) \cite{Geerits2023}
\begin{equation}\label{AFT}
    \psi^{\ell}(r,z) = \frac{1}{2\pi}\int_0^{2\pi}\mathrm{d}\phi \ \psi e^{-\mathrm{i}\ell\phi}
\end{equation}
From this we can derive the OAM distribution function
\begin{equation}\label{Dist}
    p[\ell]=2\pi \int \mathrm{d}r \mathrm{d}z \ r |\psi^\ell|^2
\end{equation}
which tells us the probability of finding the wavefunction $\psi$ in the $\ell^{\mathrm{th}}$ mode when a projective measurement of the OAM is performed. This distribution function is related to the OAM expectation value by
\begin{equation}
    \braket{\hat{L}_k}=\sum_\ell \ell \hbar p[\ell]
\end{equation}
These are the main tools we will use to analyze the OAM of neutron waves that are diffracted from perfect crystals with a non-negligible Schwinger term in the crystal potential.
\par
In quantum mechanics one can differentiate between so called intrinsic and extrinsic OAM. The term intrinsic implies that the wave OAM does not depend on the choice of reference frame \cite{Berry1998,Neil2002,Bliokh2015,Geerits2024}. The Eigenstates of the OAM operator, also referred to as pure vortex states, are an example of intrinsic OAM states. While intrinsic OAM is uniquely quantum, extrinsic OAM is not necessarily not quantum. As we showed in a previous paper \cite{Geerits2024} a superposition of two pure vortex states with neighboring mode numbers (i.e. $\ell$ and $\ell\pm 1$), carries some extrinsic OAM. While the OAM expectation value is invariant under translation, the OAM distribution function is not invariant under translation. This means that a state which can be described as a pure vortex state in one frame of reference, must be described using superpositions of vortex states in another frame of reference.
\par
Finally we introduce the concept of a linear OAM state. This is a state that exists in a superposition of two vortex modes $\ell$ and $-\ell$
\begin{equation}
    \psi(r,\phi,z)=\psi(r,z)[e^{\mathrm{i}\ell\phi}+e^{-\mathrm{i}\ell(\phi+\beta)}]
\end{equation}
with $\beta$ some arbitrary phase. The definition is analogous to linear polarization in optics which can be described as a superposition of left and right rotating polarizations. While such a state carries net-zero OAM according to the OAM expectation value, it does carry within itself two different vortex modes, which are predicted to interact differently with matter than the pure $\ell=0$ mode \cite{Afanasev2018,Afanasev2019,Afanasev2021,Jach2022}. As a result, producing such states can still be useful to investigate neutron OAM, despite the fact that the total OAM of the state is zero.
\section{Model}
Treatment of Schwinger scattering in dynamical diffraction has been covered numerous times in literature. Here we follow a simplified approach of the models described in \cite{Blume1964,Stassis1974}.
The potential inside of the crystal seen by the neutron is modeled using a strong nuclear part and an electromagnetic part, the Schwinger potential
\begin{equation}
    V=\frac{2\pi\hbar^2}{m}\sum_j b_j \delta(r-r_j)+\frac{\mu}{mc^2} \mathbf{\sigma}\cdot[\mathbf{p}\times\mathbf{E}(\mathbf{r})]
\end{equation}
with $b_j$ the coherent neutron scattering length, $\mu$ the neutron magnetic moment and $\mathbf{\sigma}$ the Pauli spin matrices. The Fourier components of the potential are then given by
\begin{equation}
    V(\mathbf{H},\mathbf{K})=\frac{2\pi\hbar^2}{m}\sum_i (b_i-2\mathrm{i} \gamma_i \mathbf{\sigma}\cdot\frac{\mathbf{K}\times\mathbf{H}}{|H|^2})e^{\mathrm{i}\mathbf{H}\cdot\mathbf{r}}
\end{equation}
\begin{figure*}
	\centering
	\includegraphics[width=1\textwidth]{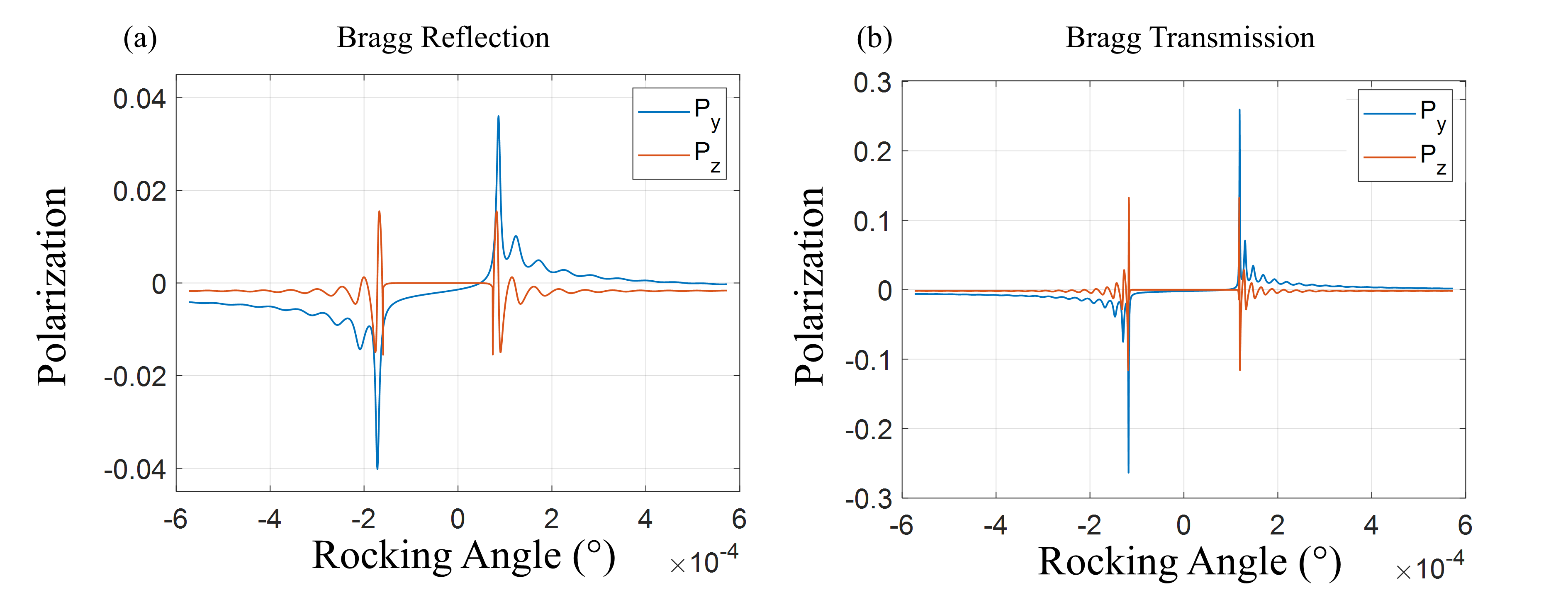}\caption{Plots of the $y$ (blue) $z$ (red) components of the neutron polarization against rocking angle after being (a) Bragg reflected or (b) transmitted from a  100 $\mu$m thick piece of quartz ([110] plane). The incident wave is initially polarized along the x direction which is chosen to be parallel to the incident wavevector. The incident wave has a wavelength of 2 $\mathrm{\AA}$}
	\label{BraggPol}
\end{figure*}
with $\gamma_i=\frac{\mu e}{\hbar c} Z_i(1-f_i(\mathbf{H}))$, $\mathbf{H}$, the reciprocal lattice vector, $\mathbf{K}$, the neutron wavevector, $Z_i$ the atomic number of the ith nucleus and $f_i$ the electronic form factor.
According to the two wave approximation in the dynamical theory of diffraction two reflected and two transmitted waves are excited within a crystal which form a pair of standing waves \cite{Batterman1964,Sears1978}
\begin{equation}\label{internalwave}
    \begin{aligned}
       &\psi^\pm_1=e^{\mathrm{i}\mathbf{k}^\pm_1\cdot\mathbf{r}}[u^\pm_1(\mathbf{0})+u^\pm_1(\mathbf{H})e^{\mathrm{i}\mathbf{H}\cdot\mathbf{r}}] \\
        &\psi^\pm_2=e^{\mathrm{i}\mathbf{k}^\pm_2\cdot\mathbf{r}}[u^\pm_2(\mathbf{0})+u^\pm_2(\mathbf{H})e^{\mathrm{i}\mathbf{H}\cdot\mathbf{r}}]
    \end{aligned}
\end{equation}
where $u^\pm_{1,2}$ represent the amplitudes of the transmitted, $u^\pm_{1,2}(\mathbf{0})$, and reflected waves, $u^\pm_{1,2}(\mathbf{H})$, with spin aligned, $+$, or anti-aligned, $-$, with the quantization axis defined by the momentum vector $\mathbf{k}_0$. The wavevectors in the crystal are given by
\begin{equation}
\begin{aligned}
&\mathbf{k}^{\pm}_{1,2}=\mathbf{k}_0+\frac{|\mathbf{k}_0|}{\cos(\gamma)}\epsilon^\pm_{1,2}\mathbf{n}
\end{aligned}
\end{equation}
with $k_0$ the incident wavevector, $\gamma$ the angle between surface normal $\mathbf{n}$ and the incident wavevector and $\epsilon^\pm_{1,2}$ (more precisely defined in the supplement) defines the spin dependent deviation of the wavevector in the crystal with respect to $k_0$, due to the periodic crystal potential (i.e. $k=k_0(1+\epsilon)$). It can be shown that the forward and reflected amplitudes are related to each other (see supplement) by the following relation 
\begin{equation}
    u_{1,2}(\mathbf{H})=-V^{-1}(\mathbf{H})[2E\epsilon_{1,2}+V(\mathbf{0})]u_{1,2}(\mathbf{0})=X_{1,2}u_{1,2}(\mathbf{0})
\end{equation}
When discussing observables, such as reflectivity, transmissivity, polarization and OAM it is more useful to look at the reflected and transmitted wavefunctions 
\begin{equation}\label{RefTransGen}
    \begin{aligned}
        &\psi_\pm^\mathbf{0}=[u_\pm^1(\mathbf{0})e^{\frac{|\mathbf{k}_0|}{\cos(\gamma)}\epsilon_\pm^{1} \mathbf{n}\cdot\mathbf{r}}+u_\pm^2(\mathbf{0})e^{\frac{|\mathbf{k}_0|}{\cos(\gamma)}\epsilon^\pm_{2} \mathbf{n}\cdot\mathbf{r}}]e^{\mathrm{i}\mathbf{k}_0\cdot\mathbf{r}}\\
        &\psi_\pm^\mathbf{H}=[u_\pm^1(\mathbf{H})e^{\frac{|\mathbf{k}_0|}{\cos(\gamma)}\epsilon_\pm^{1} \mathbf{n}\cdot\mathbf{r}}+u_\pm^2(\mathbf{H})e^{\frac{|\mathbf{k}_0|}{\cos(\gamma)}\epsilon^\pm_{2} \mathbf{n}\cdot\mathbf{r}}]e^{\mathrm{i}(\mathbf{k}_0+\mathbf{H})\cdot\mathbf{r}}
    \end{aligned}
\end{equation}
To further explore these states we must determine the amplitudes $u_{1,2}$, by determining the boundary conditions. Regardless of the geometry, the first condition is always given by the fact that at the crystal surface $\mathbf{r}\cdot\mathbf{n}$ the sum of the forward amplitudes must be equal to the incident amplitude 
\begin{equation}\label{UBC}
    {u}^\pm_1(\mathbf{0})+u^\pm_2(\mathbf{0})=u_0^\pm
\end{equation}
This is our universal boundary condition. The second boundary condition depends on the diffraction geometry (Bragg or Laue). We will start by considering the Bragg case, followed by the Laue case. In both sections we will calculate for an incident planewave the transmitted and reflected waves, including the respective intensities and polarizations as a function of $\rho$ and the rocking angle, $\theta$ (see figure \ref{CrystalAxes}). Note that the rocking angle is defined as the deviation from the Bragg angle.
From here on out all calculations will focus on the specific case of the [110] planes in quartz. As this was also used in our experiment. Furthermore quartz, especially the [110] plane, has been used in most prior experiments exploring neutron spin-orbit coupling in non-centrosymmetric crystals \cite{Forte1989,Voronin2000,Fedorov1992,Fedorov2009}.
\subsection{Bragg Case}
\begin{figure*}
	\centering
	\includegraphics[width=1\textwidth]{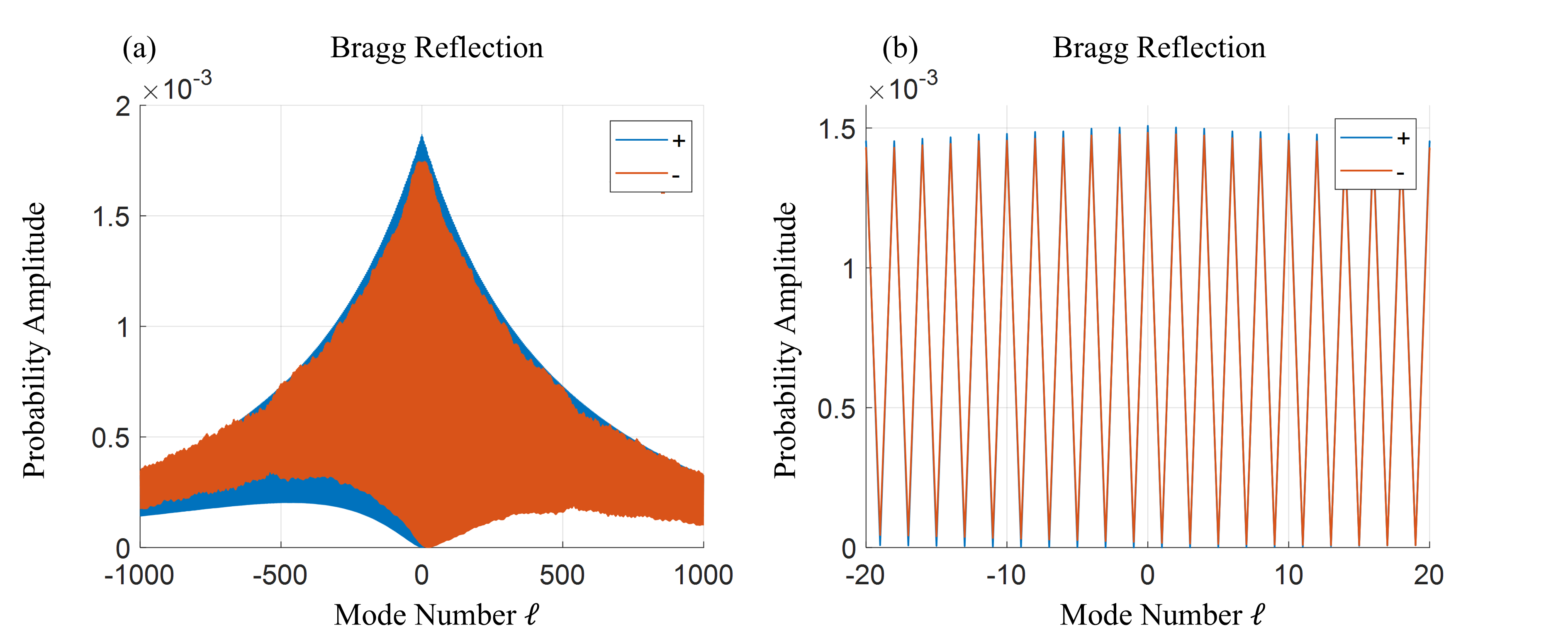}\caption{OAM distribution functions from Eq.  \ref{Dist} calculated for a 2 $\mathrm{\AA}$ neutron diffracted from a 300 micron thick piece of [110] quartz. Inset (a) shows the distribution functions of the reflected wave, while inset (b) shows a zoomed in variant. Both insets assume that $\phi$ is defined around the reciprocal lattice vector $\mathbf{H}$. The blue curve shows the non-spin flipped distribution while the red curve refers to the spin flipped distribution function. Finally we note the fast oscillations visible in (b) indicating strong suppression of neighboring modes, which in turn indicates an intrinsic OAM component.}
	\label{BraggOAMDist}
\end{figure*}
In Bragg geometry, the reflected wave exits the crystal at $\mathbf{r}\cdot\mathbf{n}=0$, while for the transmitted beam this occurs at the rear of the crystal $\mathbf{r}\cdot\mathbf{n}=D$, with $D$ the crystal thickness. Hence in Bragg geometry we can write the transmitted and reflected waves as
\begin{equation}\label{RefTrans}
    \begin{aligned}
        &\psi_\pm^\mathbf{0}=[u_\pm^1(\mathbf{0})e^{\frac{|\mathbf{k}_0|}{\cos(\gamma)}\epsilon_\pm^{1} D}+u_\pm^2(\mathbf{0})e^{\frac{|\mathbf{k}_0|}{\cos(\gamma)}\epsilon_\pm^{2} D}]e^{\mathrm{i}\mathbf{k}_0\cdot\mathbf{r}}\\
        &\psi_\pm^\mathbf{H}=[u_\pm^1(\mathbf{H})+u^\pm_2(\mathbf{H})]e^{\mathrm{i}(\mathbf{k}_0+\mathbf{H})\cdot\mathbf{r}}
    \end{aligned}
\end{equation}
From this description we can also extract the second boundary condition, which allows us to unambiguously determine the amplitudes. This condition states that there is no reflected wave field at the rear end of the crystal
\begin{equation}
    \psi_\pm^\mathbf{H}(\mathbf{r}\cdot\mathbf{n})=[u_\pm^1(\mathbf{H})+u^\pm_2(\mathbf{H})]=0
\end{equation}
From this condition and Eq. \ref{UBC}, we can derive the amplitudes for the Bragg case
\begin{equation} \label{BraggAmplitudes}
\begin{aligned}
    &u_1({\mathbf{0}})=-(E_1X_1-E_2X_2)^{-1}E_2X_2 u_0 \\
    &u_2({\mathbf{0}})=-(E_2X_2-E_1X_1)^{-1}E_1X_1 u_0
\end{aligned}
\end{equation}
\begin{figure*}
\includegraphics[width=1\textwidth]{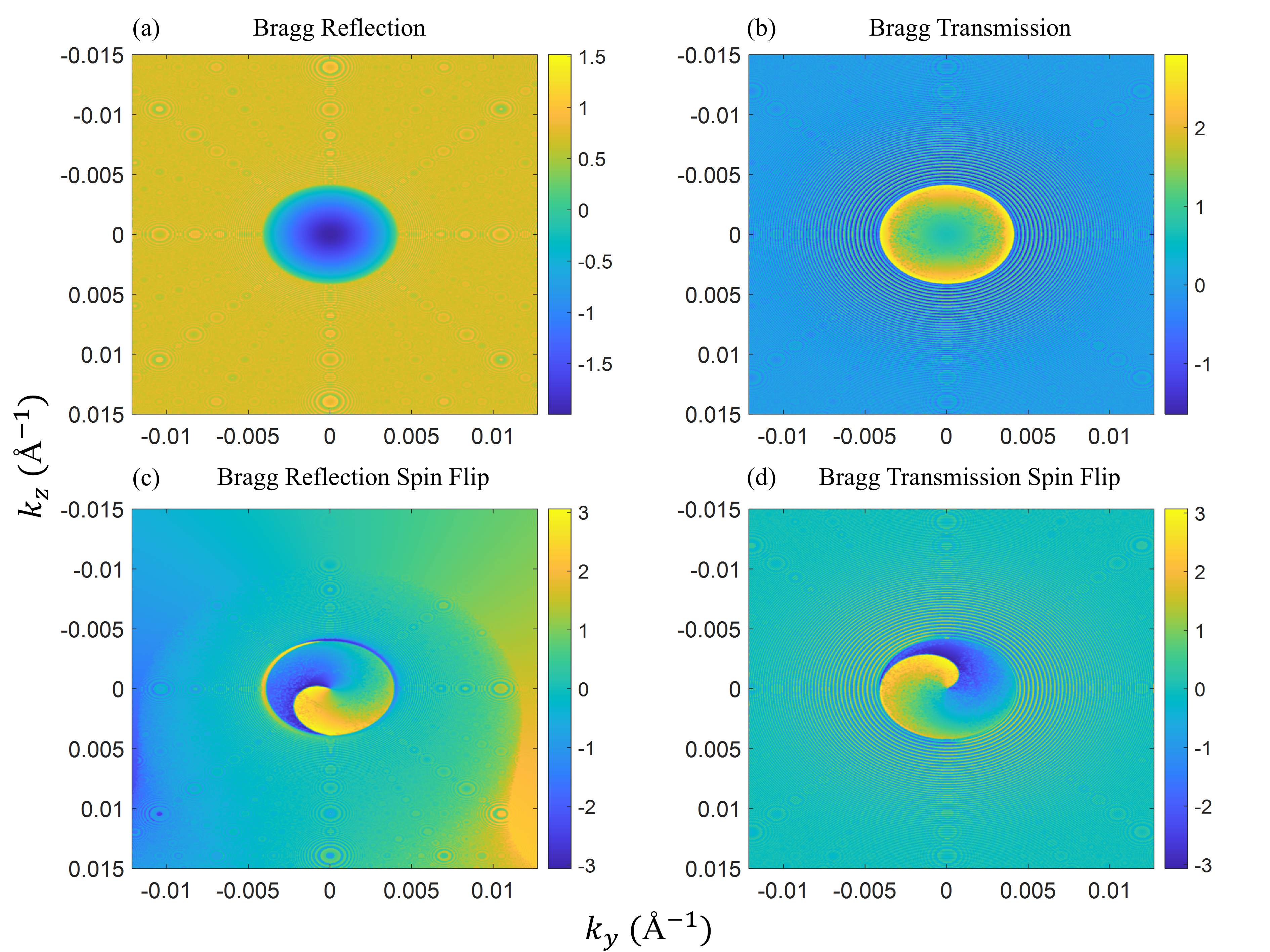}\caption{Phase of the (a) reflected and (b) transmitted non spin flipped wavefunctions and the (c) reflected and (d) transmitted spin flipped wavefunctions (eq. \ref{RefTrans}). The latter show a clear vortex structure "spinning" in opposite directions. The vortex line is the line on which the phase flips from $+\pi$ to $-\pi$. Since there is only one vortex line (or curve) in this case we can conclude that the mode number is equal to $\ell=\pm 1$. The incident wave has a wavelength of 5.0279 $\mathrm{\AA}$ with a spin polarized along the momentum direction.}
	\label{BraggVortices}
\end{figure*}
\begin{figure*}
	\centering
	\includegraphics[width=1\textwidth]{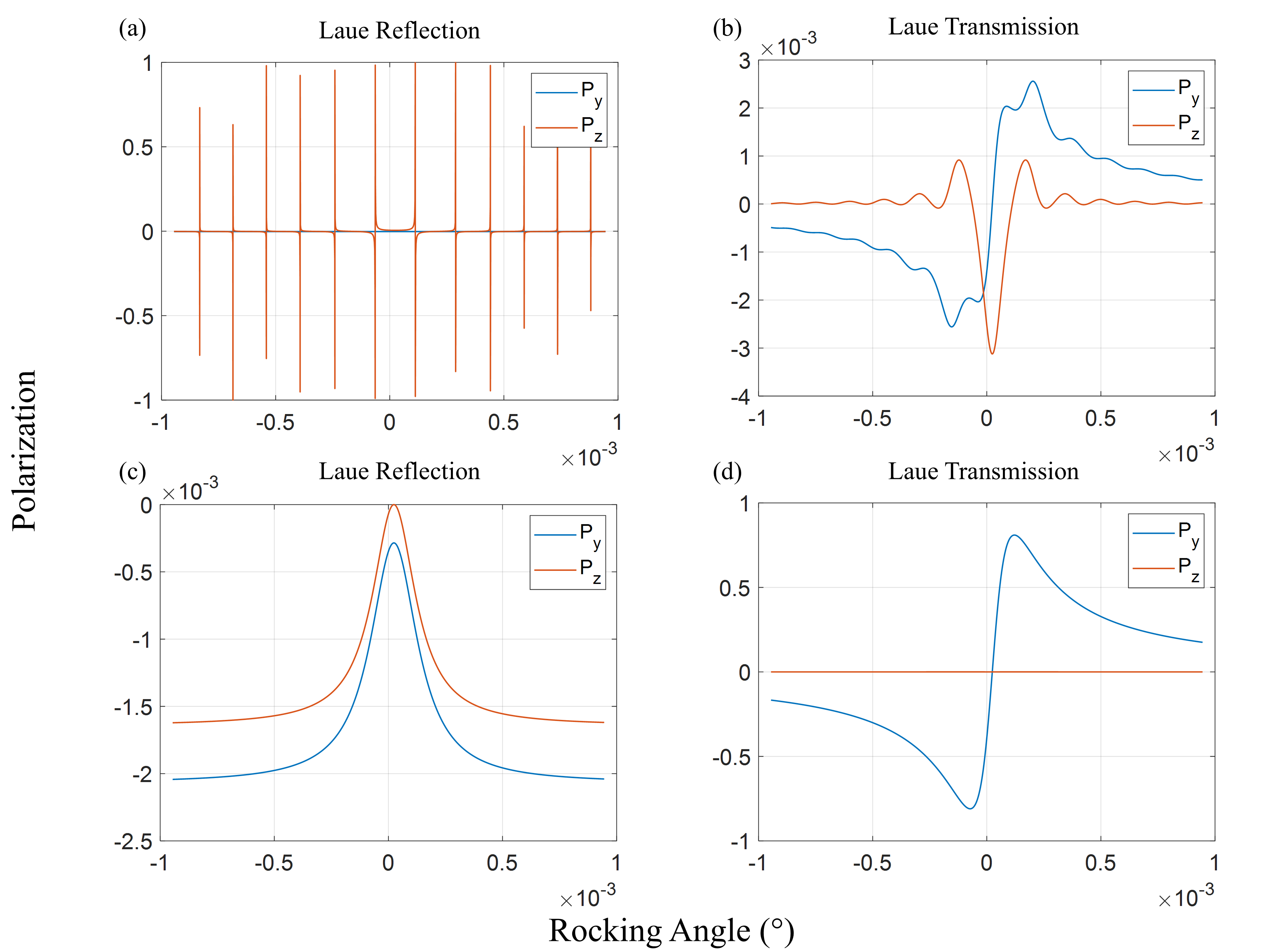}\caption{Plots of the $y$ (blue) $z$ (red) components of the neutron polarization against rocking angle after being Laue transmitted (eq. \ref{LauePolEq}) from a (a) 100 $\mu$m and (c) 35 mm thick piece of quartz ([110] plane). Insets (b) and (d) show the polarization components of the reflected waves for both thickness respectively. The incident wave is initially polarized along the $x$ direction which is chosen to be parallel to the incident wavevector. The incident wave has a wavelength of 2 $\mathrm{\AA}$. For the thicker crystal an approximation is used which averages over the Pendelösung oscillations (visible in the top insets), this aids in visibility of the polarization.}
	\label{LauePol}
\end{figure*}
with $E_{1,2}=e^{\frac{|\mathbf{k}_0|}{\cos(\gamma)}\epsilon_\pm^{1,2} D}$
Equations \ref{BraggAmplitudes} and \ref{RefTrans} are sufficient to determine the polarization and OAM of the reflected and transmitted wavefunctions. In previous work where we examine OAM generation in the Schwinger interaction \cite{Geerits2021}, we demonstrate that OAM is generated in the spin flipped state and the total angular momentum of the neutron is conserved. Basing ourselves on these findings we will investigate the OAM distribution of the spin flipped components of the reflected and transmitted wavefunctions (Eq. \ref{RefTrans}). \par Assuming an incident beam initially polarized along the momentum direction (also denoted here as the $x$-direction) we will first calculate the polarization of the  $\mathbf{P}_R$ and transmitted $\mathbf{P}_T$ beams
\begin{equation}\label{LauePolEq}
    \begin{aligned}
&\mathbf{P}_R=\frac{<\psi_R(\mathbf{r}\cdot\mathbf{n}=0)|\mathbf{\sigma}|\psi_R(\mathbf{r}\cdot\mathbf{n}=0)>}{R}\\	&\mathbf{P}_T=\frac{<\psi_T(\mathbf{r}\cdot\mathbf{n}=D)|\mathbf{\sigma}|\psi_T(\mathbf{r}\cdot\mathbf{n}=D)>}{T}
    \end{aligned}
\end{equation}
with $R=<\psi_R(\mathbf{r}\cdot\mathbf{n}=0)|\psi_R(\mathbf{r}\cdot\mathbf{n}=0)>$ the reflectivity and $T=<\psi_T(\mathbf{r}\cdot\mathbf{n}=D)|\psi_T(\mathbf{r}\cdot\mathbf{n}=D)>$ the transmission. Figure \ref{BraggPol}, shows the polarization component, orthogonal to the incident polarization, of the Bragg reflected and transmitted waves.
For both reflected and transmitted beams the $y$-polarization component flips, when the rocking angle and therefore the transverse momentum component is flipped. As a result neutron spin and neutron momentum are coupled or entangled \cite{Shen2020,Hasegawa2010}. \par
In addition we argue that this $y$-polarization component is evidence that the spin flipped wavefunction in both the reflected and transmitted case carries OAM with respect to the non-spin flipped wavefunction. This stems from the fact that the orthogonal polarization components are basically the real and imaginary parts of the interference between the spin flipped and non spin flipped wavefunctions. The flipping of the amplitude seen in $P_y$, which occurs when the momentum vector is mirrored around the reciprocal lattice vector, is indicative that the amplitude of the spin flipped wavefunction also flips with respect to the non-spin flipped wavefunction when this transverse momentum component is inverted. \par
To definitively demonstrate that the spin-flipped wavefunctions carry some OAM we must analyze the two dimensional transverse structure of the wavefunctions. Meaning we must numerically calculate $\psi_\pm^{\mathbf{0},\mathbf{H}}$ at the crystal surfaces as a function of rocking angle $\theta$ and the tilting angle $\rho$, which are related to the transverse wavevectors by $k_y=|k_0|\theta$ and $k_z=|k_0|\rho$. Using $\psi_\pm^{\mathbf{0}}(\mathbf{r}\cdot\mathbf{n}=D,k_y,k_z)$ and $\psi_\pm^{\mathbf{0}}(\mathbf{r}\cdot\mathbf{n}=D,k_y,k_z)$ we can calculate the OAM distribution functions (Eq. \ref{Dist}) of the spin flipped and spin-non-flipped wavefunctions at the crystal surface. Results are shown in figure \ref{BraggOAMDist}
These distribution functions demonstrate that the spin flipped wave function gains some OAM with respect to the non-spin flipped wavefunction. One can also see that both OAM distributions of the spin flipped wavefunction and non-spin flipped wavefunction oscillate (i.e. the odd modes are suppressed), indicating linear OAM states. It should be noted that these OAM distribution functions depend strongly on the $k_y-k_z$ domain over which they are calculated. Here we used a realistic domain where the $\theta$ range is limited by the rocking width and the $\rho$ range is determined by a typical beams divergence, 1 degree. \par
In our previous paper on OAM generation in electric fields \cite{Geerits2021} we demonstrate that pure intrinsic longitudinal OAM states are only generated if the momentum vector and electric field vector are parallel. As a result the probably most useful diffraction configuration is a backscattering mode, as in this case the effective electric field of the crystal is parallel to the incident momentum vector. Figure \ref{BraggVortices}, shows the phase structure of the spin flipped and non spin flipped wavefunctions reflected and transmitted from [110] quartz in a back scattering configuration. This corresponds to a wavelength of 5.0279 $\mathrm{\AA}$, very close to the peak of a cold neutron spectrum.
It is immediately evident from the phase structure of the spin flipped wavefunctions that these possess an OAM of $\ell=1$ or $\ell=-1$, while the non-spin flipped wavefunctions carry no OAM, consistent with our findings in \cite{Geerits2021}. We conclude that backscattering from a non-centrosymmetric crystal in Bragg geometry acts as an OAM raising/lowering operator coupled to a spin flip operator, thereby entangling the spin and OAM degrees of freedom. The additional detail that the OAM carrying part of the wavefunction is also spin flipped, makes it exceptionally easy to detect as a spin filter could be used to separate the spin flipped beam from the non-scattered beam, even in a backscattering geometry. The limiting factor with this method is the low efficiency. For a 10 mm thick quartz crystal the ratio between the integrated spin flipped reflectivity and the non spin flipped reflectivity is on the order of $10^{-6}$. The same is true for the transmitted wave. Note that these ratios improve for shorter wavelengths (6\% at 2 $\mathrm{\AA}$), however as shown above the spin-orbit states produced with thermal neutrons are not pure. We will find in Laue geometry the efficiency is much higher, even in backscattering, since the neutrons must propagate through the whole crystal.

\subsection{Laue Case}
\begin{figure*}
	\centering	\includegraphics[width=1\textwidth]{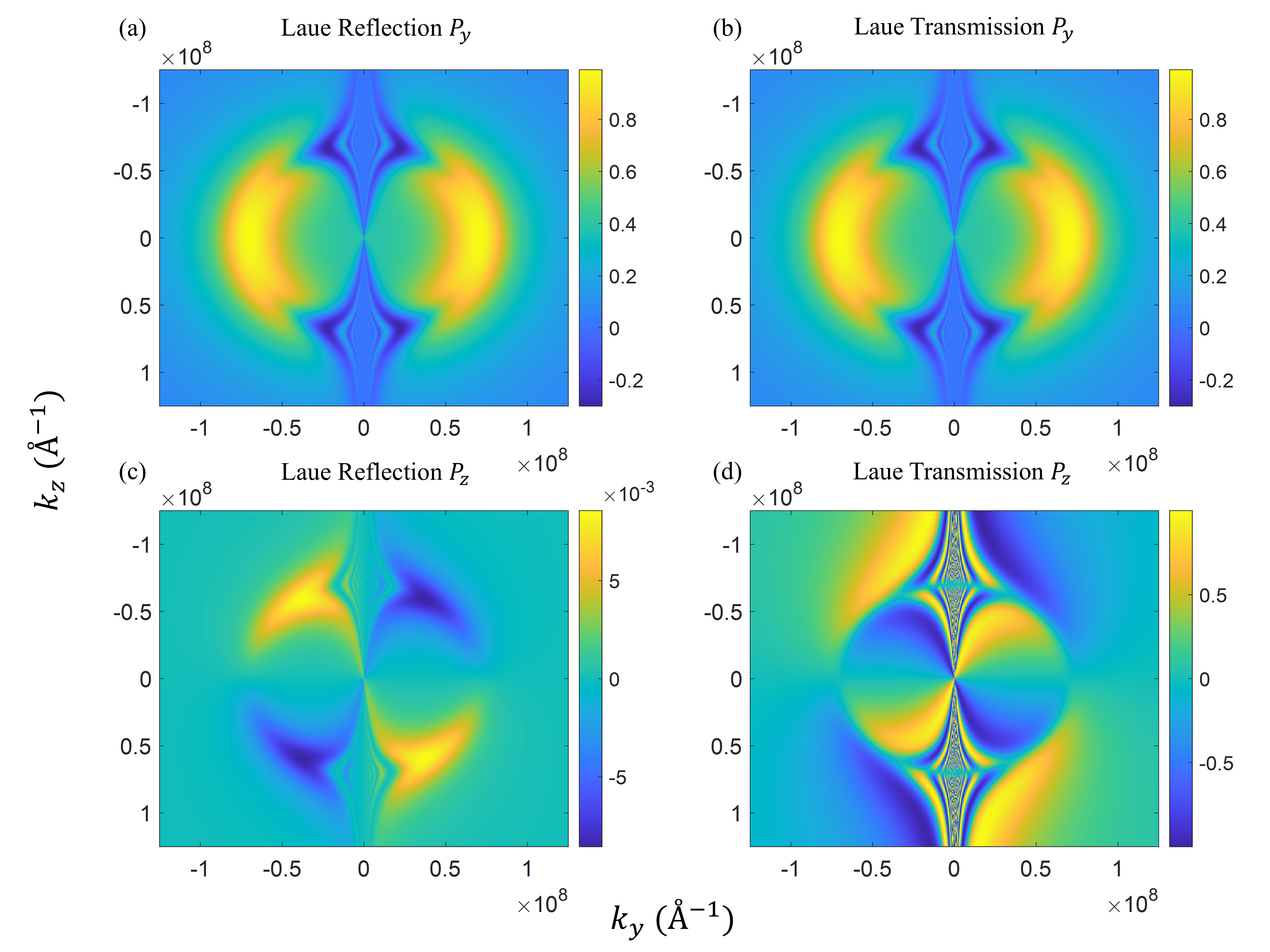}\caption{Orthogonal polarization components of a neutron diffracted from a 35 mm thick piece of [110] quartz according to equation \ref{LauePolEq}. The initial polarization is chosen along the momentum direction $x$. The insets each show a different polarization component (a) $P_y$ and (b) $P_z$ of the reflected beam and (c) $P_y$ and (d) $P_z$ of the transmitted beam. Since the wavelength of the incident beam is chosen such that we are in a Backscattering configuration, $\lambda=5.0279 \mathrm{\AA}$, $k_z$ going through 0 represents the rocking angle surpassing 90 degrees which cannot technically be accomplished in a true Laue geometry, unless the entry surface of the crystal is tilted. Note the colorbar to the left of each figure, indicating that $P_y$ of the transmitted and reflected waves are almost identical, however $P_z$ of the reflected wave is a few orders of magnitude weaker than that of the transmitted wave.}
	\label{LauePol2}
\end{figure*}
We now proceed to look at vortex state generation by diffraction from [110] quartz in Laue geometry. In this geometry both the reflected and transmitted waves leave the crystal at $\mathbf{r}\cdot\mathbf{n}=D$. Thus the wave functions are given by
\begin{equation}\label{RefTransLaue}
    \begin{aligned}
        &\psi_\pm^\mathbf{0}=[u_\pm^1(\mathbf{0})e^{\frac{|\mathbf{k}_0|}{\cos(\gamma)}\epsilon_\pm^{1} D}+u_\pm^2(\mathbf{0})e^{\frac{|\mathbf{k}_0|}{\cos(\gamma)}\epsilon_\pm^{2} D}]e^{\mathrm{i}\mathbf{k}_0\cdot\mathbf{r}}\\
        &\psi_\pm^\mathbf{H}=[u_\pm^1(\mathbf{H})+u^\pm_2(\mathbf{H})]e^{\mathrm{i}(\mathbf{k}_0+\mathbf{H})\cdot\mathbf{r}}
    \end{aligned}
\end{equation}
Hence we find the second boundary condition with is unique to Laue geometry. This condition states that there is no reflected wavefield at the crystal entrance, leading to the following expression
\begin{equation}	       X_1u_1(\mathbf{0})+X_2u_2(\mathbf{0})=0
\end{equation}
Together with our universal boundary condition Eq. \ref{UBC} we can derive the spinor amplitudes of the Laue beams
\begin{equation}
	\begin{aligned}
		u_1(\mathbf{0})=(X_2-X_1)^{-1}X_2u_0\\
		u_2(\mathbf{0})=-(X_2-X_1)^{-1}X_1u_0
	\end{aligned}
\end{equation}
Just as before when looking at the Bragg case we will first determine the polarization of the reflected and transmitted waves 
\begin{equation}
	\begin{aligned}
		\mathbf{P}_R=\frac{<\psi_R(\mathbf{r}\cdot\mathbf{n}=D)|\mathbf{\sigma}|\psi_R(\mathbf{r}\cdot\mathbf{n}=D)>}{R}\\
		\mathbf{P}_T=\frac{<\psi_T(\mathbf{r}\cdot\mathbf{n}=D)|\mathbf{\sigma}|\psi_T(\mathbf{r}\cdot\mathbf{n}=D)>}{T}
	\end{aligned}
\end{equation}
Assuming the incident beam is spin polarized longitudinally we show the orthogonal polarization components of the reflected/transmitted beams against rocking angle in figure \ref{LauePol}. The incident wave is assumed to be thermal with a wavelength of $2 \mathrm{\AA}$. Similar to the Bragg case the Laue transmitted polarizations indicate spin-orbit entanglement, as the spin flips under a parity inversion of the momentum vector. We note that the spin rotation can be much stronger in the Laue case, since neutrons can propagate through the full length of large crystals. Curiously, the reflected beam does not display the same polarization signature indicating spin-orbit entanglement.  
\begin{figure*}
	\centering	\includegraphics[width=1\textwidth]{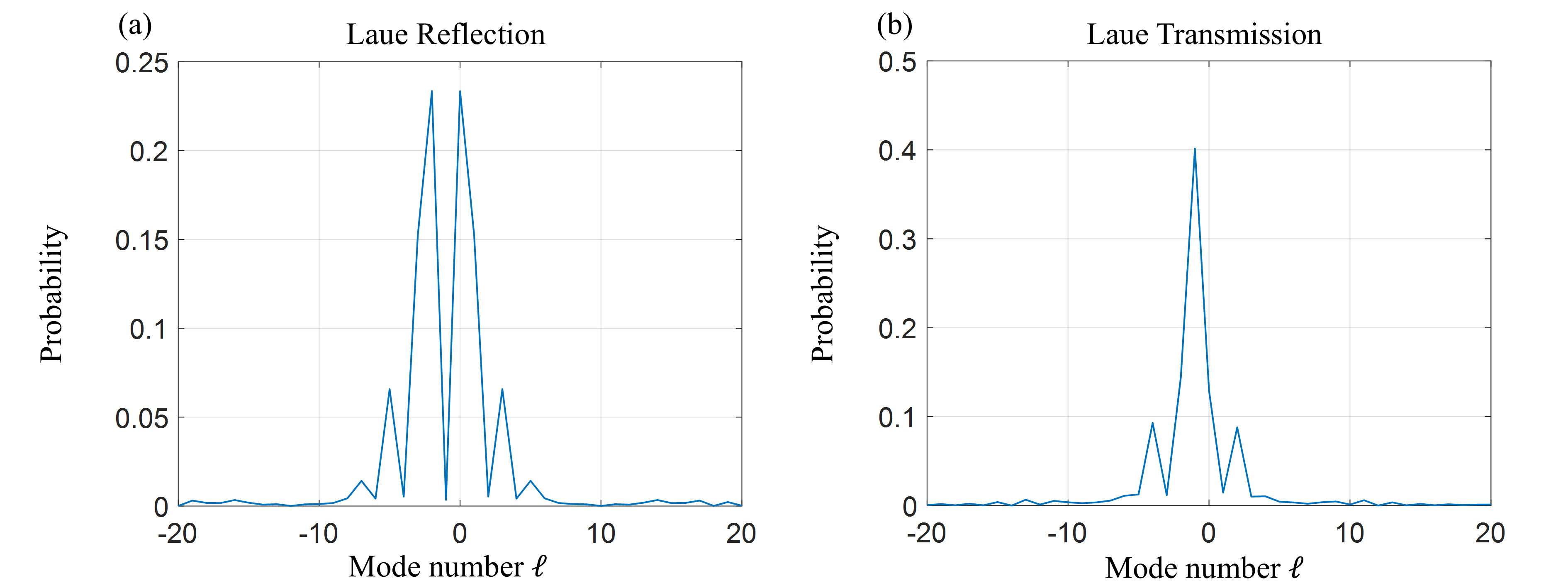}\caption{OAM distribution functions of the product between the conjugate of the up spin wavefunction with the down spin wavefunction, $\psi_+^*\psi_-$. Inset (a) shows the reflected wave case, while inset (b) looks at the transmitted wave. Both waves OAM distributions have an average mode number of $\bar{\ell}=-1$.}
	\label{LaueDistInt}
\end{figure*}
When we delve deeper into the matter we find that the OAM states produced in Laue diffraction aren't as "clean" as those produced in Bragg diffraction. In the next step the transmitted wavefunction produced in back diffraction will be calculated numerically, assuming that the initial spin is polarized along the flight direction. In the lab Laue back diffraction is usually not possible, since the angle between $\mathbf{n}$ and $\mathbf{k}_0$ would be 90 degrees and the incoming wave would therefore not impact the crystal surface. One could circumvent this limitation as we could tilt the surface slightly to allow for a Bragg angle close to 90 degrees. In the following calculations we will however assume a flat surface and look at Bragg angles close to 90 degrees.  Figure \ref{LauePol2} shows the orthogonal polarization components of the reflected and transmitted back-diffracted beam, assuming the incident wave was longitudinally polarized. The [110] quartz crystal has a length of $35$ mm.
The polarization is divided into two lobes, the left lobe represents the physically reachable side of the rocking curve below 90 degree Bragg angle, while the right lobe goes beyond 90 degrees. In both lobes we see that the transverse polarization components of the transmitted wave follow very closely the azimuthal coordinate vector $\mathbf{\hat{\phi}}$, indicating a coupling between polarization and orbit (i.e. spin-orbit entanglement).
As expected from our previous analysis the reflected beam exhibits a much weaker coupling. \par
\begin{figure*}
	\centering
	\includegraphics[width=1\textwidth]{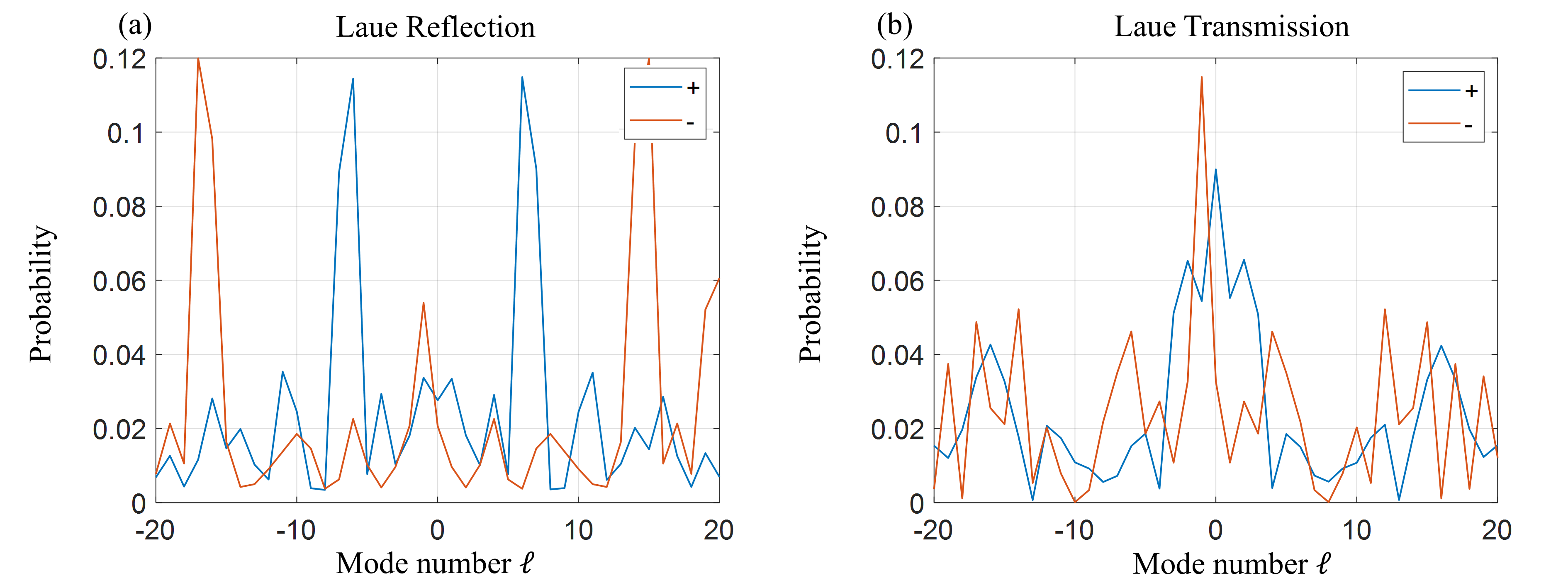}\caption{
    OAM distribution functions of the spin flipped wavefunction $\psi_-$ (red) and non spin flipped wavefunction $\psi_+$ (blue) for the (a) transmitted and (b) reflected case.
    The average mode number of the spin flipped wavefunctions is $\bar{\ell}=-1$, while the non spin flipped wavefunctions have an average mode number of zero.}
	\label{LaueOAMDist}
\end{figure*}

Orthogonal polarization components are proportional to the interference between the non spin flipped $\psi_+$ and spin flipped $\psi_-$ states. A polarization vortex therefore indicates that the OAM of the up and down spin states are different. So, to further explore this we will calculate the OAM distribution function (Eq. \ref{Dist}) of $\psi_+^*\psi_-$. By using this product common phase factors between the up and down wavefunctions can be eliminated. These distribution functions are shown in figure \ref{LaueDistInt}. Note that from this point forward all analysis focuses on the left lobe of the wavefunctions indicated in figure \ref{LauePol2}, as this area can experimentally be reached.
We see that for both the reflected as well as the transmitted beam the OAM distribution of the interference term $\psi_+^*\psi_-$ is centered on $\ell=-1$ indicating that indeed the spin flipped wavefunction carries one unit of OAM with respect to the non-spin flipped wavefunction. Neither of these vortices are as "clean" as in the Bragg case, though the transmitted beam is closer to a pure vortex than the reflected beam. We can finally look at the OAM distribution functions calculated using the pure wavefunctions $\psi_+$ and $\psi_-$. These are shown in figure \ref{LaueOAMDist}.
Once again we find that the spin flipped wavefunction has its OAM changed by one unit of $\hbar$ with respect to the non-spin flipped wavefunction. However the OAM distribution functions also display symmetric sidebands, which arise from Pendellösung oscillations in the wavefunction. These are proportional to $\cos(kx)=\cos(kr\cos(\phi))$, which when plugged into the azimuthal Fourier transform produce sidebands. Contrary to the Bragg case significant spin flip scattering takes place in Laue geometry, hence this geometry can be used to efficiently convert spin to OAM.

We may conclude that both Bragg and Laue diffraction can be used to produce OAM states in thermal neutrons. OAM produced by Bragg diffraction is cleaner and has a higher purity, however only in the Laue case can a sufficient intensity of OAM carrying neutrons be produced. A 35 mm thick crystal is sufficient to produce a maximally entangled spin-orbit state (i.e. a $\pi/2$ pulse). This observation leads us to conclude that experiments such as those described in \cite{Forte1989} and \cite{Voronin2000}, unknowingly produced twisted spin-orbit neutron waves before the field of twisted neutrons was even established.
\section{Experiment}
\begin{figure*}
	\centering
	\includegraphics[width=1\textwidth]{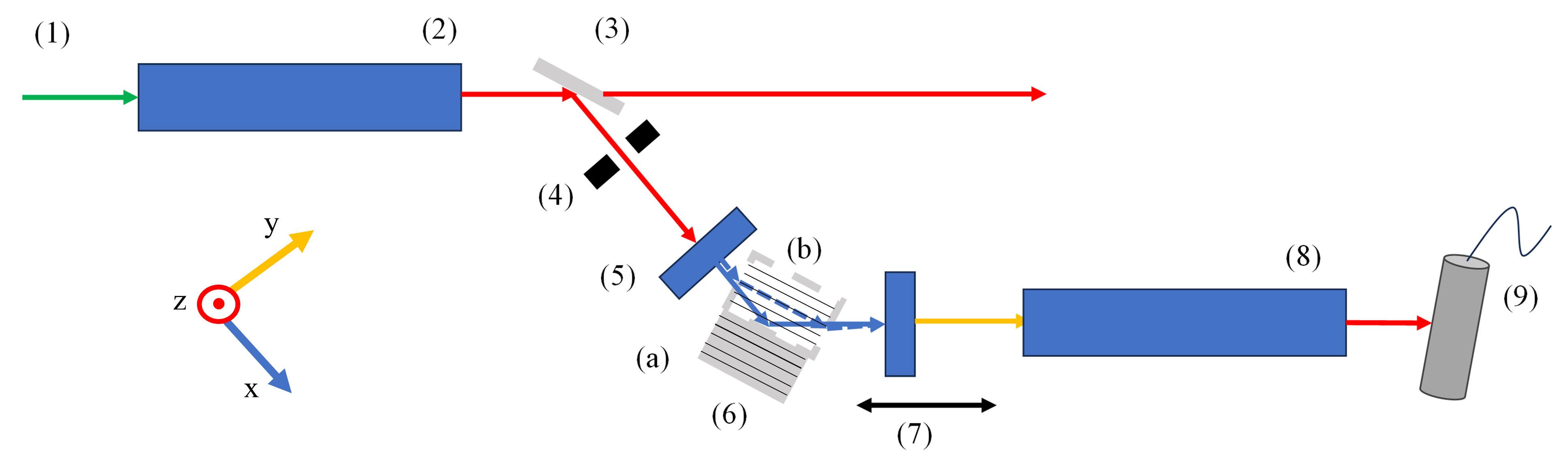}\caption{Schematic of the setup used to test spin-orbit state generation in neutrons by dynamical diffraction. A broad-band unpolarized (green) neutron beam comes from the guide (1) and enters a bender mirror (2) where the beam is polarized along $z$ (red). A perfect crystal quartz monochromator (3) reflects a small portion of the beam into the setup. The crystal is cut such that the [110] planes are parallel to the crystal surface. The Bragg angle is chosen to be 22.5 degrees which results in a wavelength of 1.8 $\mathrm{\AA}$, close to the peak of the spectrum. Neutrons continue from the monochromator through a slit collimator (4) to a $\pi/2$ flipper, which prepares the neutron spin along the $x$ direction (blue). After this the neutron is Bragg (a) or Laue (b) reflected from a second "test" quartz crystal (6), cut in the same orientation as the monochromator. A movable $\pi/2$ flipper (7) paired with a polarization analyzer (8) project the neutron spin on the $y$ direction (yellow). Finally the neutrons are detected using a He3 counter (9). A 1 mT guide field surrounds the entire setup. The test crystal is attached to a pair of rotation stages, enabling rotations around the $z$-axis (rocking) and the $x$-axis (tilt).}
	\label{DelftSetup}
\end{figure*}

We now present a neutron optical experiment testing orbital angular momentum generation in Schwinger diffraction from the [110] plane of perfect quartz. Experiments described here were carried out at a thermal test beamline of the 2.3 MW reactor of the Reactor Institute Delft. Our experiment investigates polarization patterns produced in Bragg and Laue reflection of thermal (wavelength: $1.8$ $\mathrm{\AA}$) neutrons. The setup used for these experiments is shown in figure \ref{DelftSetup}. 
A bender polarizer is inserted in the neutron guide. Polarized neutrons exiting the polarizer then, in the guide, encounter a quartz crystal (25x25x5 $\mathrm{mm}^3$) cut parallel to the [110] plane. The angle of the crystal with respect to the beam is selected to maximize the reflected flux (corresponding to wavelength of $1.8$ $\mathrm{\AA}$). Reflected neutrons pass through a slit and a Mezei flipper which prepares the spin along the momentum vector. After spin preparation the neutrons reflect from a 25x25x35 $\mathrm{mm}^3$ (height x width x length) test quartz crystal, also cut parallel to the [110] planes. The crystal is mounted on a series of stages allowing for translation to switch between Bragg and Laue geometries. In addition the stages enable rocking and tilting of the crystal. Once reflected neutrons pass through a second Mezei flipper on a translation stage and a polarization analyzer. The combination of flipper and analyzer allows us to analyze the neutron polarization along all three axes. Finally neutrons are detected in a Helium-3 proportional counter. We note that a cadmium slit could be scanned vertically in front of the detector to measure polarization as a function of vertical divergence angle.
Measuring in transmission was not possible at the time due to higher background and noise considerations. We will begin by presenting the results of the Bragg reflection measurements for which we expect a non zero result (see figure \ref{BraggPol}), followed by Laue reflection, where our theory predicts a null result (see figure \ref{LauePol}). 
\subsection{Bragg Geometry}
\begin{figure*}
	\centering
	\includegraphics[width=1\textwidth]{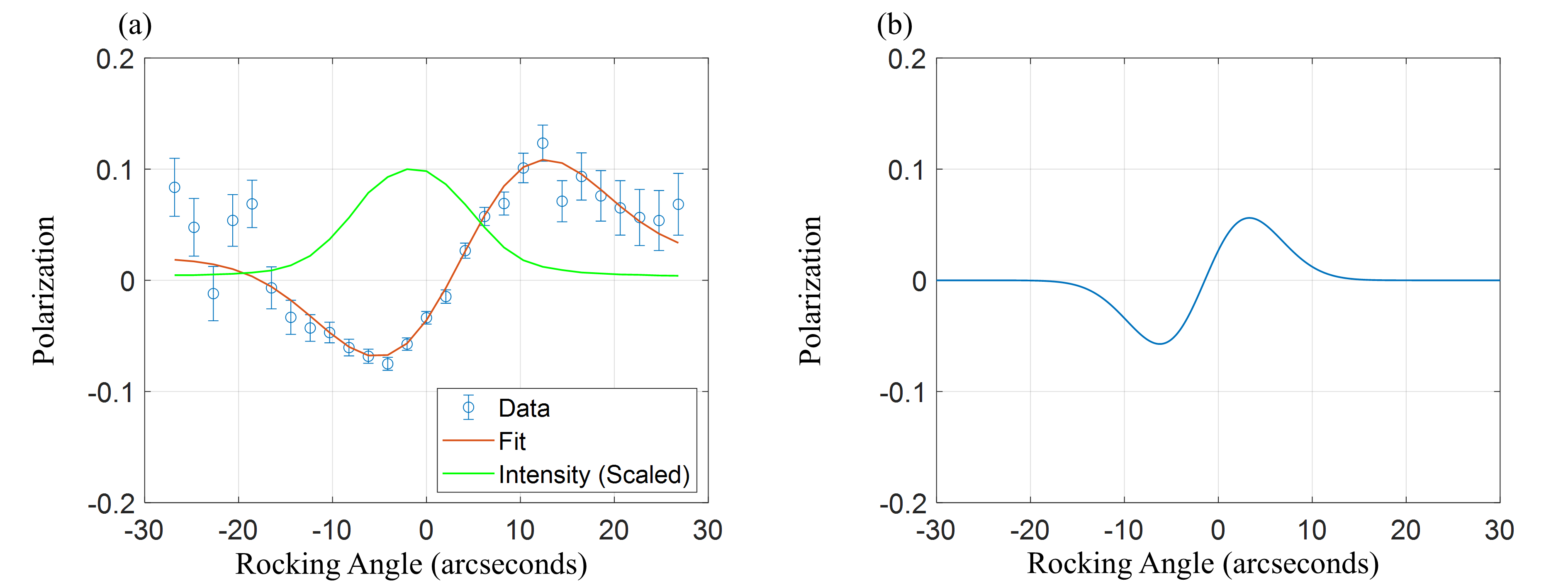}\caption{(a) Plots of the $y$ polarization component measured against rocking angle in arcseconds. The data is shown in blue, while a fit, based on the first derivative of a Gaussian is shown in red. The rocking curve is shown scaled in green. This measurement is compared to (b) a model which convolves the expected polarization calculated using dynamical theory of diffraction with a Gaussian resolution function, to account for the momentum spread of the monochromator.}
	\label{ExperimentRockingPolarization}
\end{figure*}

\begin{figure*}
	
	\includegraphics[width=1\textwidth]{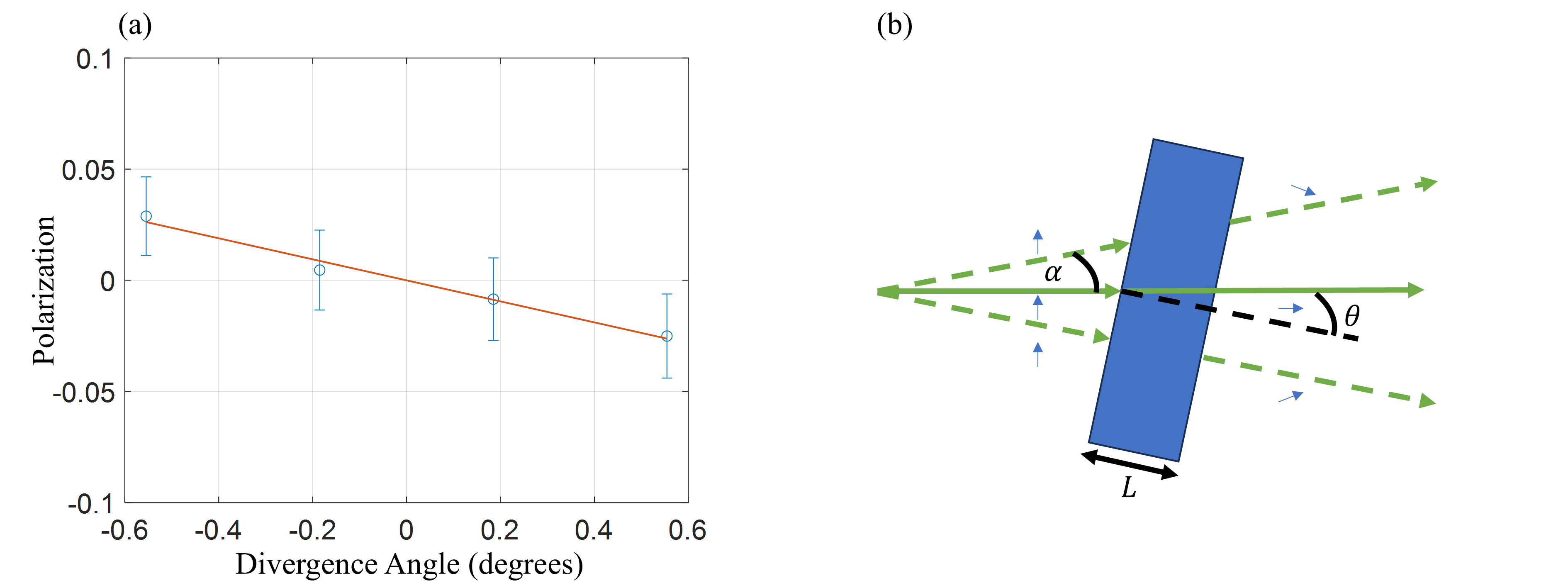}\caption{(a) Plot of $z$ (vertical) polarization component measured as a function of vertical divergence in degrees. A linear fit is shown in red. (b) Hypothetical mechanism by which this momentum-spin coupling is produced. A divergent neutron beam (green) passes through a slightly tilted coil (blue square) (side view). Due to the different path lengths taken through the coil, the spin (blue arrows) rotation angle becomes coupled to the divergence angle.}
	\label{VerticalDivergenceCoupling}
\end{figure*}

We begin by looking at the Bragg case. The rocking curve width of our test crystal appears about five times larger than the theoretical estimate of 1 asec, obtained from the dynamical theory of diffraction, for the [110] planes of quartz, indicating some imperfection in our crystals (mechanical stress, surface defects, mosaicity) or a mismatch between the tilt angles of the monochromator and test crystal. In prior studies perfect quartz crystals have occasionally shown such stress/strain effects leading to a gradient in the lattice spacing \cite{Fedorov2009}. Regardless of what caused the broadening, a washing out of the rocking curve will also lead to a washing out of the coupling. 
Our experiment measured, intensity against rocking angle and vertical divergence as well as the $y$ and $z$ polarization components, the former against rocking angle and the latter against vertical divergence angle. 
The $y$-polarization as well as a scaled rocking curve is shown in figure \ref{ExperimentRockingPolarization} (a) together with a simple model which convolves the calculated polarization (i.e. figure \ref{BraggPol}) with the estimated resolution function of the monochromator, here we use a Gaussian, in inset b. Qualitatively the two agree in terms of form. Quantitatively, slight differences can be seen in terms of amplitude and width, likely due to the fact that the resolution function of the monochromator is not exactly known, in addition to possible defects in the test crystal (described above), which perturb the polarization. Nonetheless, the data shows evidence for neutron spin neutron orbit coupling as the polarization rotates by 180 degrees for a 180 degree rotation of the transverse momentum. Since the dynamical theory does not predict a significant coupling between any other momentum component with any other polarization component for thermal neutrons it is tempting to conclude that we are dealing with a linearly polarized OAM state (i.e. a superposition of $\ell=\pm m$). Even so we examined the other polarization components as well to further characterize the spin-orbit state. To this end the z-polarization was measured against divergence in z-direction. This is similar to tilting the crystal (i.e. changing the $\rho$ angle). Recall in the case of an $\ell=-1$ or $\ell=+1$ state this polarization component ought to mimic the shape of the $y$-polarization against rocking angle (i.e. a flip in the divergence angle should flip the polarization component). Again the dynamical theory of diffraction predicts no such coupling for thermal neutrons in quartz, however other spin-optical components in the setup may introduce such a spin-momentum dependence. Our measurement was implemented by vertically scanning a thin horizontal cadmium slit between the analyzer and detector and measuring the $z$-polarization against the position of the slit. The analyzer coil (nr. 7 in figure \ref{DelftSetup}) was disabled for this measurement. The polarization against the estimated vertical divergence angle is shown in figure \ref{VerticalDivergenceCoupling}.
Despite the large error, the data shows a clear linear dependence of the z-polarization on the vertical divergence angle. Our model using dynamical diffraction predicts a spin rotation on the order of $10^{-3}$ radians, smaller than the measurement error. Hence a different mechanism is necessary to explain this coupling. Such a mechanism is proposed in figure \ref{VerticalDivergenceCoupling} (b). If the preparation coil is tilted even by a small amount, neutrons traversing through the coil with a divergence in direction of the coil tilt experience a shorter path length and therefore less spin rotation than neutrons traveling straight through the coil. In addition neutrons diverging under an angle opposite the tilt angle of the coil experience an even larger path length causing the spin to be over-rotated. This divergence dependent phase can be expressed as
\begin{equation}
	\delta\phi=\gamma B \frac{\Delta L}{v} =\frac{\pi}{2}[\frac{1}{\cos(\theta)}-\frac{1}{\cos(\theta-\alpha)}]
\end{equation}
$\theta$ is the tilt angle of the coil and $\alpha$ the divergence angle of the beam. In the derivation of the above formula we have assumed that B is chosen such that for $\alpha=0$ the coil provides a $\pi/2$ flip. For small $\theta$ this equation is second order in $\alpha$. In the case of a 5 degree coil tilt the spin phase would change by $5\cdot10^{-4}$ radians over a 1 degree divergence change. However, if an external guide field is present a larger B field is required to produce the $\pi/2$ leading to a larger divergent dependent phase. For a 1 mT guide field and a 5 degree coil tilt the spin phase as a function of divergence increases significantly to $1.5\cdot10^{-2}$. This method of tilting a magnetic field region to generate spin-momentum coupling, here used unintentionally, is the basic principle behind Spin Echo Small Angle Neutron Scattering (SESANS) \cite{Pynn1978,Rekveldt1996} and the coherent averaging method for producing OAM states \cite{Sarenac2018,Sarenac2018b,Sarenac2019,Geerits2024_2}. Our experiment demonstrates the feasibility of combining the Schwinger and coherent averaging methods to produce OAM states in thermal neutrons. 
\par 
However figure \ref{VerticalDivergenceCoupling} demonstrates that $P_z$ is rather small compared to $P_y$ shown in figure \ref{ExperimentRockingPolarization}, hence on the momentum scale of the rocking width the OAM state of the wavefunction can be understood as a linearly polarized state. Adding a sufficiently strong vertical magnetic gradient however, could produce an $\ell=1$ or $\ell=-1$ state on the same length/momentum scale. Note that this length/momentum scale is quite important, as for thermal neutrons diffracted from quartz the vertical momentum spread is much wider than in rocking direction, hence when calculating the expectation value integral $<L_z>$ the OAM washes out when integrated over the whole space. If one however constrains integration to a length/momentum scale similar to that of the rocking width one would find a non-zero value, even in the thermal neutron case.
\subsection{Laue Geometry}
\begin{figure}[!b]
	\centering
	\includegraphics[width=0.5\textwidth]{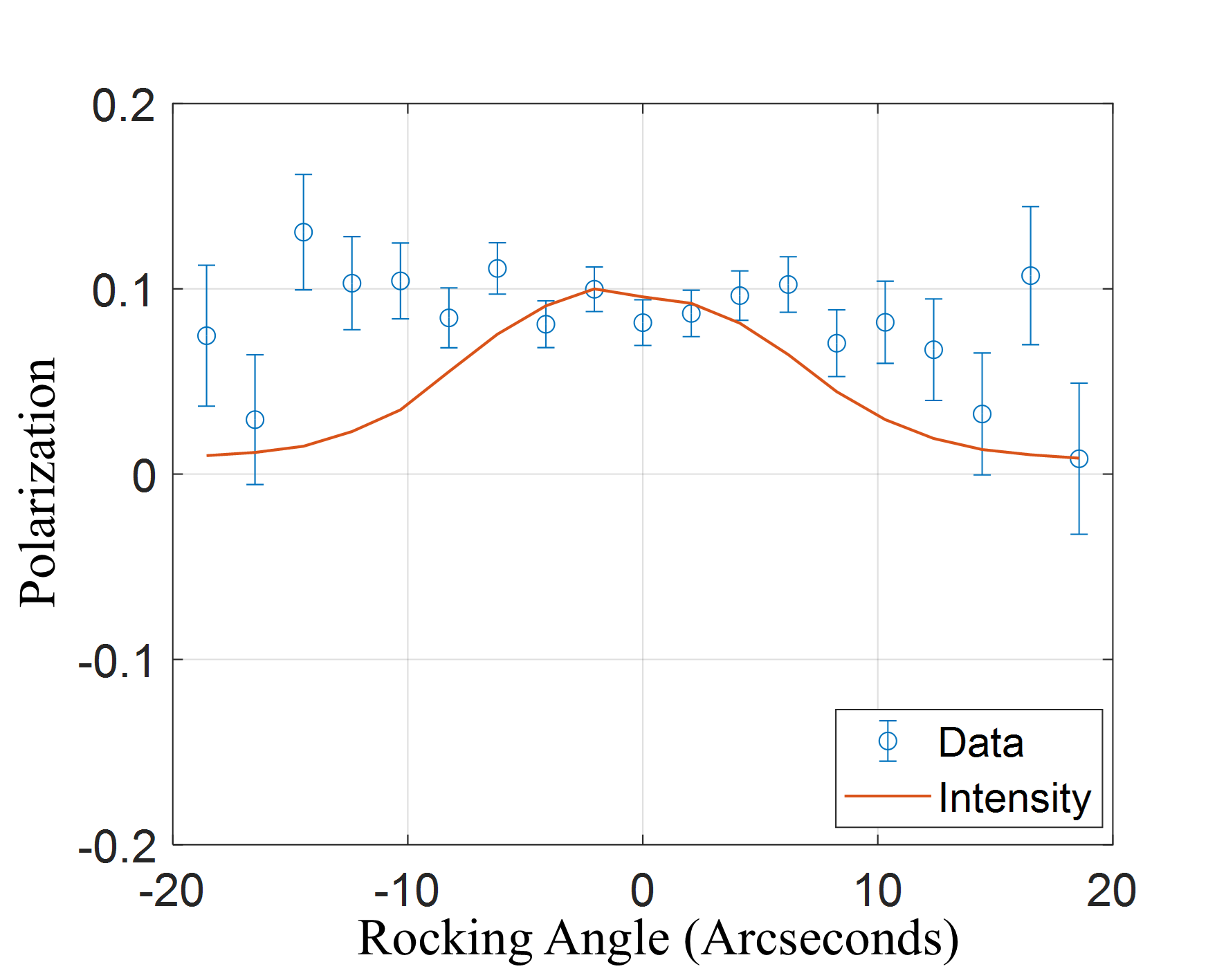}\caption{Plot of the $y$-component of the neutron polarization against rocking angle in Laue reflection geometry. The data is shown in blue, while the intensity/rocking curve is shown in red.}
	\label{LauePolExperiment}
\end{figure}
We now move on to Laue geometry where the polarization of the reflected beam was measured as a function of rocking angle. Recall that the theory of dynamical diffraction predicts no visible coupling between momentum and polarization in this case. In this measurement the crystal was moved into Laue geometry ((b) in figure \ref{DelftSetup}). Again the preparation coil was used to prepare the neutron spin along the momentum direction, while the analysis coil together with the polarization analyser measured the orthogonal, y component of the polarization. This y component is measured against rocking angle and shown in figure \ref{LauePolExperiment}. 
Our measurement confirms the lack of polarization rotation predicted by the dynamical theory of diffraction. However we may observe that the measured rocking width is much wider than the predicted width according to the dynamical theory, hence any Schwinger effect would also be weakened. \par\noindent Recall that our theory does predict that the spin flipped wavefunction carries OAM, however the observable evidence for this isn't as obvious as in the Laue transmission or Bragg case, where we can observe a coupling between polarization and momentum.
\section{Discussion}
In 2021 it was demonstrated that strong electric fields can act as an intermediary, allowing neutron spin to OAM conversion \cite{Geerits2021}. Though for this process to be efficient fields would have to be extraordinarily large. It has been well known that non centrosymmetric crystals can present large effective electric fields to neutrons \cite{Forte1989,Fedorov1992,Alexeev1989}. Here we have shown using the theory of dynamical diffraction that these fields can be exploited to facilitate highly efficient spin to OAM conversion. Especially in the case of back diffraction (i.e. a rocking angle close to 90 degrees) highly pure OAM states can be produced. While this is especially true in the Bragg case, we demonstrated that the spin flip probability for back-diffraction is practically non-existent. As a result we conclude that only Laue back-diffraction, in particular in transmission mode, can realistically generate pure intrinsic longitudinal vortex states in an efficient way. As a result we conclude that earlier experiments which used cold neutrons being Laue diffracted from perfect quartz at Bragg angles close to 90 degrees \cite{Voronin2000,Fedorov2009} may have unknowingly produced spin-orbit entangled beams
\begin{equation}
    \ket{\psi_i}=\ket{\ell=0}\ket{\uparrow} \rightarrow \ket{\psi_0}=\frac{1}{\sqrt{2}}(\ket{\ell=0}\ket{\uparrow}+\ket{\ell=1}\ket{\downarrow})
\end{equation}
the depolarization effects observed in these experiments can then simply be understood as a consequence of the orthogonality of the $\ell=0$ and $\ell=1$ states. \par
Since our method is only feasible in back-diffraction it follows that it is a monochromatic technique. Furthermore the requirement of perfect crystals limits the usable flux. Though if we compare our approach with the only other known method of producing vortex beams, diffraction from forked gratings \cite{Sarenac2022}, we find that the technique described here may have some advantages. Due to considerations of transversal coherence and required optical path lengths the fork gratings described in \cite{Sarenac2022} operate at the tail end of the cold neutron spectrum ($12\mathrm{\AA}$) and owing to the short optical path length of the gratings, they are still far from maximal efficiency. Our method shares some of these issues, arising from the narrow bandwidth diffracted by perfect crystals, however this technique can be operated close to the peak of the cold flux, thereby greatly increasing the number of OAM neutrons available. In addition our method supports a higher conversion efficiency for $\ell=0\rightarrow|\ell|=1$, coupled to the fact that this technique does not simultaneously produce states of opposite parity. On the flip side this perfect crystal approach requires a polarized beam and cannot raise/lower the OAM of a beam by multiple units of $\hbar$. As a result the choice of technique will depend on the desired application. \par
In addition to the theoretical work we have presented an experiment confirming our results for thermal neutrons Bragg and Laue reflected from perfect quartz. Coupled to our model we showed that in thermal neutrons diffraction from non-centrosymmetric crystals can produce linear OAM states (i.e. superpositions of $\ell=m$ and $\ell=-m$), which despite having an OAM expectation value of zero, have been shown to behave differently than pure $\ell=0$ states \cite{Geerits2024} and have also been predicted to have different scattering and absorption properties \cite{Afanasev2019,Jach2022}. Furthermore we have made a plausible case that our technique could create pure vortex states when combined with coherent averaging \cite{Sarenac2018b,Sarenac2019,Geerits2023}. This has the potential of simplifying OAM generating optics for thermal neutrons. In addition to produce thermal vortex neutrons one could use pairs of crossed crystals or use higher order reflections in back-diffraction geometry, though both of these techniques significantly impact the available flux, hence it is likely that for thermal neutrons real space coherent averaging would like be the best approach to produce vortex states \cite{Geerits2023,Geerits2024_2}. \par
Despite this we believe that our technique will pave the way for producing relatively pure vortex neutrons at the peak of the cold flux, thereby enabling neutron optical experiments investigating OAM dependent cross sections \cite{Afanasev2018,Afanasev2019,Afanasev2021,Jach2022} and quantum information experiments with four degrees of freedom, especially useful for contextuality \cite{Shen2020} and weak measurement type experiments such as the Cheshire cat \cite{Danner2024}.
\section{Acknowledgments}
N.G. A-S. Berger and S.S. acknowledge funding from the Austrian Science Fund (FWF), Project No. P34239. N.G. is also supported by the US Department of Energy (DOE) Grant No. DE-SC0023695. The authors express gratitude for beam time granted by the Reactor Institute Delft in 2022, for conducting the experiments reported on in this paper.  
\pagebreak
\section{References}
\bibliographystyle{unsrt}
\bibliography{SpinOrbitBibliography}
\end{document}